\documentclass[sigconf]{acmart}
\setcopyright{none}
\usepackage[T1]{fontenc}
\usepackage[utf8]{inputenc}
\usepackage[english]{babel}
\usepackage{acronym}
\usepackage{longtable}
\usepackage{textcomp}
\usepackage{multirow}
\usepackage{makecell}
\usepackage{graphicx}
\usepackage{color}
\usepackage{epsfig}
\usepackage{url}
\usepackage{float}
\usepackage{array}
\usepackage{verbatim}
\usepackage{pifont}
\usepackage{tabularx}
\usepackage{hyperref}
\usepackage{comment}
\usepackage{caption}
\usepackage{algorithm2e}
\usepackage{listings}

\AtBeginDocument{%
  }

\newcolumntype{P}[1]{>{\centering\arraybackslash}p{#1}}

\acrodef{AP}{Access Point}
\acrodef{API}{Application Program Interface}
\acrodef{IC}{Integrated Circuit}
\acrodef{BLE}{Bluetooth Low Energy}
\acrodef{CA}{Certification Authority}
\acrodef{SDR}{Software Defined Radio}
\acrodef{DoS}{Denial of Service}
\acrodef{DH}{Diffie Hellman}
\acrodef{ECC}{Elliptic Curve Cryptography}
\acrodef{ECDH}{Elliptic Curve Diffie Hellman}
\acrodef{GNSS}{Global Navigation Satellite System}
\acrodef{GPS}{Global Positioning System}
\acrodef{IoT}{Internet of Things}
\acrodef{KDF}{Key Derivation Function}
\acrodef{HKDF}{HMAC Key Derivation Function}
\acrodef{MEO}{Medium Earth Orbit}
\acrodef{MITM}{Man-in-the-Middle}
\acrodef{PKC}{Public Key Cryptography}
\acrodef{PKI}{Public Key Infrastructure}
\acrodef{RF}{Radio Frequency}
\acrodef{RFID}{Radio Frequency Identification}
\acrodef{RSS}{Received Signal Strength}
\acrodef{SNR}{Signal-to-Noise-Ratio}
\acrodef{TLS}{Transport Layer Security}
\acrodef{PEPP-PT}{Pan-European Privacy-Preserving Proximity Tracing}
\acrodef{TCN}{Temporary Contact Numbers}
\acrodef{UHF}{Ultra High Frequency}
\acrodef{TDMA}{Time Division Multiple Access}
\acrodef{GFSK}{Gaussian Frequency Shift Keying}
\acrodef{DPSK}{Differential Phase Shift Keying}
\acrodef{FHSS}{Frequency-Hopping Spread Spectrum}
\acrodef{PACT}{Private Automated Contact Tracing}
\acrodef{RSS}{Received Signal Strength}
\acrodef{RSSI}{Received Signal Strength Indicator}
\acrodef{DP-3T}{Decentralized Privacy-Preserving Proximity Tracing}
\acrodef{LTE}{Long Term Evolution}
\acrodef{LZMA}{Lempel–Ziv–Markov chain Algorithm}
\acrodef{EXT}{Extended file system}
\acrodef{YAFFS}{Yet Another Flash File System}
\acrodef{JFFS2}{Journalling Flash File System version 2}
\acrodef{RTOS}{Real-Time Operating System}
\acrodef{RAM}{Random-Access Memory}
\acrodef{JTAG}{Joint Test Action Group}

\graphicspath{{images/}}

\acmYear{2024}
\acmConference[ ]{}
\acmBooktitle{ }
\acmDOI{10.1145/3664476.3664511}

\begin{document}

\title{Reverse Engineered MiniFS File System}

\author{Dmitrii Belimov}
\affiliation{%
  \institution{Technology Innovation Institute}
  \city{Abu Dhabi}
  \country{United Arab Emirates}}
\email{dmitrii.belimov@tii.ae}

\author{Evgenii Vinogradov}
\affiliation{%
  \institution{Technology Innovation Institute}
  \city{Abu Dhabi}
  \country{United Arab Emirates}}
\email{evgenii.vinogradov@tii.ae}

\renewcommand{\shortauthors}{Belimov et al.}

\begin{abstract}
In an era where digital connectivity is increasingly foundational to daily life, the security of Wi-Fi Access Points (APs) is a critical concern. This paper addresses the vulnerabilities inherent in Wi-Fi APs, with a particular focus on those using proprietary file systems like MiniFS found in TP-Link's AC1900 WiFi router. Through reverse engineering, we unravel the structure and operation of MiniFS, marking a significant advancement in our understanding of this previously opaque file system. Our investigation reveals not only the architecture of MiniFS but also identifies several private keys and underscores a concerning lack of cryptographic protection. These findings point to broader security vulnerabilities, emphasizing the risks of security-by-obscurity practices in an interconnected environment. Our contributions are twofold: firstly, based, on the file system structure, we develop a methodology for the extraction and analysis of MiniFS, facilitating the identification and mitigation of potential vulnerabilities. Secondly, our work lays the groundwork for further research into WiFi APs' security, particularly those running on similar proprietary systems. By highlighting the critical need for transparency and community engagement in firmware analysis, this study contributes to the development of more secure network devices, thus enhancing the overall security posture of digital infrastructures. 
\end{abstract}


\maketitle

\section{Introduction}
\label{sec:intro}
Nowadays, the world is more connected than ever before. This development is largely enabled by the widespread availability and use of wireless networks. For example, Wi-Fi \acp{AP} enabled easy Internet access in homes, offices, and public spaces. Their ubiquity and the convenience they offer have made them critical devices for different activities such as research activities, communications, and entertainment. However, the potential vulnerabilities of these wireless \acp{AP} raised concerns from the security perspective~\cite{makhdoom2018anatomy},~\cite{Nadir2022}.

A malicious user can exploit Wi-Fi AP vulnerabilities remotely to penetrate an internal network~\cite{9074317}. Once inside, the attacker gains access to all network data and resources. Devices that adopt proprietary solutions without disclosing firmware details or security techniques are particularly vulnerable~\cite{291237}. This concern is accentuated in the case of low-end devices employing undocumented proprietary file systems like MiniFS, used in multiple TP-Link WiFi routers. The absence of publicly available information on such systems complicates efforts to understand and secure them, making proprietary file systems a focal point of security analysis. When attackers discover security vulnerabilities within the firmware, they can leverage these for executing remote attacks~\cite{romana2020security},\cite{hampton2015survey},\cite{tsow2006warkitting}.

Indeed, a more robust security paradigm is achieved when the wider community has access to and can scrutinize the firmware. Open analysis not only aids in identifying and rectifying vulnerabilities more efficiently but also promotes the development of more secure systems. Thus, the need for transparent and accessible security measures is paramount in safeguarding these devices. Our investigation into the MiniFS file system embedded within TP-Link's routers underscores this point, aiming to demystify its structure and enhance the security framework of Wi-Fi APs through detailed analysis and reverse engineering. 


\noindent
\textbf{Contributions:} Our work details the results of the reverse engineering of the MiniFS file system used in TP-Link's AC1900 WiFi router~\cite{tp_link_vendor}, leading to the successful extraction of the device's stored files. We describe the file system structure and its components. Based on the revealed structure, we develop the file system extraction flow and present preliminary results of data extraction. The presented contributions streamline the security analysis for devices using MiniFS, enabling users to identify and fix potential vulnerabilities.

\noindent
{\bf Roadmap:} In our paper, we systematically explore the security landscape of Wi-Fi access points, focusing on the MiniFS file system within the TP-Link AC1900 router. Beginning with \textbf{Section~\ref{sec:background}}, we lay the groundwork by discussing our firmware security testing methodology alongside a review of the limited literature on MiniFS. \textbf{Section~\ref{sec:materials}} details our experimental setup, including the tools for firmware extraction and analysis. In \textbf{Section~\ref{sec:minifs_struct}}, we delve into the heart of our research, presenting a thorough description of MiniFS's structure, uncovering its unique features and variances from known systems. \textbf{Section~\ref{sec:eva}} extends our exploration to the practical extraction of the file system, documenting the process and showcasing a wide array of decoded files. The paper is concluded in \textbf{Section~\ref{sec:conclusion}}, where we synthesize our discoveries, ponder their security implications, and chart potential paths for future inquiries. This narrative aims to enrich the understanding of firmware security and shed light on the enigmatic MiniFS file system.

\section{Background and Related Works}
\label{sec:background}
In this section, we outline a generic firmware security testing methodology that will be applied to analyze a WiFi router firmware. Additionally, we delve into the existing body of work surrounding the MiniFS file system. Despite its use in various low-end devices, detailed academic discussions on MiniFS are sparse. Our review aims to bridge this gap by systematizing current knowledge about this proprietary file system.

\subsection{Firmware security testing methodology}
\label{sec:steps}
The security testing of device firmware involves several critical steps, each aimed at acquiring/extracting, analyzing, and testing the firmware to identify its functionality and discover potential security threats \cite{Nadir2022,bakhshi2024review}. As depicted in Figure~\ref{fig:RE}, the process mainly defined in $6$ steps:
\begin{figure}
    \centering
    \includegraphics[width=\columnwidth]{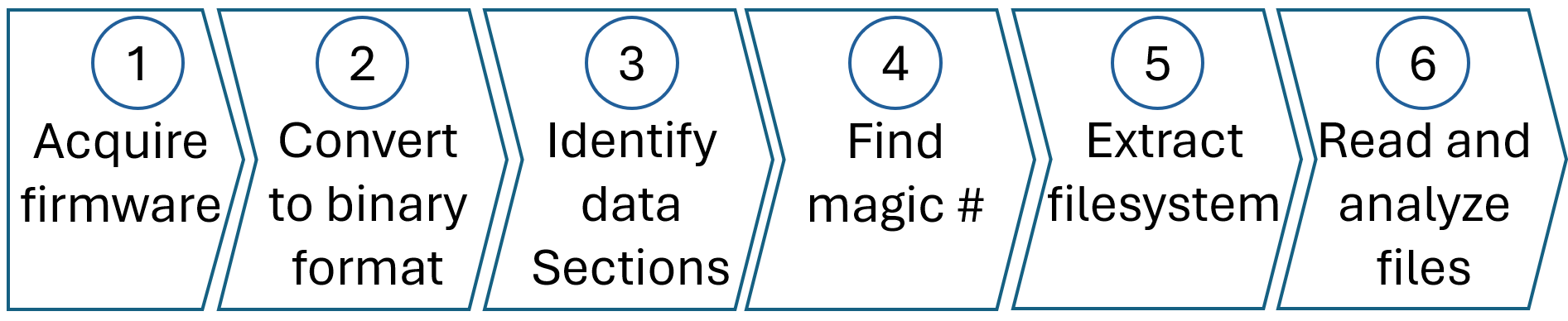}
    \caption{Firmware security testing procedure.}
    \label{fig:RE}
\end{figure}
\subsubsection{Step 1 - Firmware Acquisition}
    There are multiple options for acquiring firmware from a router/access point~\cite{vasile2019breaking}: 
    \begin{itemize}
        \item \textit{Option 1 - Download via Internet}: The firmware can be acquired via Internet by (i) downloading it from the manufacturer website, or (ii) by recovering it from the administration panel of the router.
        \item \textit{Option 2 - Eavesdropping}: Leveraging tool such as Wireshark~\cite{wireshark_link} or \texttt{tcpdump}~\cite{tcpdump_link} the firmware can be acquired by eavesdropping the communication during the software update procedure. It is worth noticing that in this case the firmware could be encrypted by the manufacturer. Further, \acp{SDR} can be adopted to intercept and demodulate the update signals to extract the firmware.
        \item \textit{Option 3 - Physical Access}: The firmware is recovered from a physical access of the device. The attacker adopts hardware tools such as \ac{JTAG} debuggers, serial adapters, and specialized software to to dump the firmware data directly from the chip~\cite{shwartz2018reverse}.
    \end{itemize}
\subsubsection{Step 2 - Convert firmware into binary format}
Depending on the options used in Step 1, the firmware dump may have different formats. For instance, when a memory dump is acquired at Step 1 following  Option 3, the reader output is a binary file. However, extracting the binary file from an update in Option 2 may require a more sophisticated approach.

\subsubsection{Step 3 - Identify data section}
Firmware often contains encrypted or compressed sections. Compressed sections might have identifiable signatures, although these are not always present. Identifying encrypted sections involves analyzing the firmware image's entropy, which can reveal both encrypted and compressed areas. Observing the entropy's rising and falling edges helps delimit these sections. Often, format signatures or algorithms used to generate them can be found at the section edges.
Moreover, the entropy value allows us to make assumptions about the state of the data: high entropy values indicate usage of data compression or encryption algorithms.

\subsubsection{Step 4 - Find Magic Numbers}
File systems often write a magic number in the header to indicate the file system type. This magic number denotes the name or acronym of the file system in ASCII code. To complete the data section identification, we need to find magic numbers serving as identifiers for file formats, embedded in the firmware binary. 

Tools such as `binwalk` \cite{binwalk} are instrumental in automating the search for magic numbers within firmware images, providing insights into the composition of the firmware and highlighting areas of interest for deeper analysis. However, if a proprietary file system is used, this tool is not applicable.

\subsubsection{Step 5 - Identify and Extract File System}
After identifying magic numbers, the next critical phase is the identification and extraction of the file system embedded within the firmware. This step is fundamental for accessing the firmware's file structure and contents. Analysts use tools like `binwalk` or custom scripts for proprietary systems to extract the file system. This process allows for the examination of file hierarchies, system configurations, and embedded applications. For proprietary file systems not readily identifiable by common tools, analysts may need to reverse-engineer the file system structure based on patterns observed in the binary data.

\subsubsection{Step 6 - Analyze Files and Identify Vulnerabilities}
With the file system extracted, the next step involves a thorough analysis of the files contained within. This includes:
\begin{itemize}
    \item \textit{Static Analysis}: Examining the code without executing it to find vulnerabilities, hardcoded credentials, or suspicious patterns \cite{chess2004static}. Tools like IDA Pro \cite{idapro_link}, Ghidra \cite{ghidra_link}, or simpler hex editors can be used for this purpose.
    \item \textit{Dynamic Analysis}: Running the firmware or its components in a controlled environment (emulation or virtualization) to observe its behavior \cite{zaddach2014avatar}. This can reveal vulnerabilities such as buffer overflows, authentication bypasses, or insecure communication protocols.
    \item \textit{Configuration and Binary File Analysis}: Reviewing configuration files, scripts, and binary executables to identify insecure settings or vulnerabilities that could be exploited by attackers. This also includes searching for backdoors, undocumented features, or potential points of entry.
\end{itemize}

The described steps represent a systematic approach ensuring a thorough examination of the firmware, uncovering any potential security issues that could be exploited. The findings from this process enhance the security of the device.

\subsection{Previous version of the MiniFS file system}
\label{sec:mfs_background}
MiniFS is a small, lightweight file system designed for use in embedded systems with limited resources. It is commonly used in WiFi routers (e.g., TP-Link and Mercusys). One of the key benefits of MiniFS is its small size. It typically requires only a few kilobytes of memory, making it ideal for use in devices with limited resources. 


Despite its utilization in various devices, to the best of our knowledge, the MiniFS file system was not described in the peer-reviewed literature. Initial contributions from several online resources \cite{minifs_article_1},\cite{minifs_article_2},\cite{GIToldMINIFS} have laid the groundwork for understanding MiniFS, albeit with limited comprehensive details. This section aims to consolidate the existing knowledge while identifying the gaps yet to be filled. 

Key aspects identified from the available resources include:

\begin{itemize}
    \item The MiniFS file system is identifiable by a unique magic string "MINIFS". 
    \item The structure of MiniFS includes a header, file property records, and data chunks.   
    \item The header of MiniFS, crucial for navigating the file system, comprises a marker followed by an unidentified field with an offset of \texttt{0x10}.    
    \item File properties within MiniFS are encoded using $88$-byte fixed-size records. These records contain the file size and its offset in the data chunk. A single 80-byte field is used to store the file name and path.
    \item Chunks contain the file data. Each chunk contains data belonging to a single file.  
    \item \ac{LZMA} compression is utilized for storing files. Additionally, the first 4 bytes of a chunk serve as the \ac{LZMA} configuration word (\texttt{0x5A000080}).
\end{itemize}

While these elements provide a foundational understanding of MiniFS, we did not manage to reproduce the flow presented in \cite{minifs_article_1},\cite{minifs_article_2},\cite{GIToldMINIFS} while analyzing our device in question (see Section~\ref{sec:device}). Consequently, we have another version of MiniFS at our disposal. In Section~\ref{sec:minifs_struct}, we describe this new version of the file system and underscore the discovered differences. 

\section{Materials}
\label{sec:materials}
The testing methodology given above is applicable to many devices. In this subsection, we describe the device under test and the tools we applied to perform the testing steps presented above.

\subsection{Device in question}
\label{sec:device}
As an example of a budget WiFi AP, we selected TP-Link AC1900 WiFi Router. It is a popular device (marked as `Hot Buys` on the manufacturer's website) owing its popularity to an affordable price. 
We have found that it used VxWorks \ac{RTOS}~\cite{10273771} and a custom private read-only MiniFS file system. 
\subsection{Tools for security testing}
\begin{figure}
    \centering
    \includegraphics[width=0.48\textwidth]{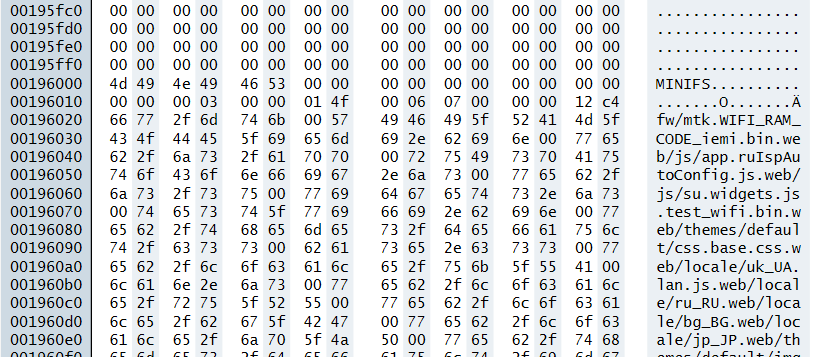}
    \caption{Raw binary data with MiniFS.}
    \label{fig:raw_no_mark}
\end{figure}

While performing the steps presented in Section~\ref{sec:steps}, we used:

\begin{itemize}
    \item \textbf{Steps 1-2:} We have desoldered the flash memory and dumped its content into a binary data file with the programmer Xgecu TL866II Plus \cite{tl866_link}.
    \item \textbf{Step 3:} Entropy Analysis has been performed with 'Binwalk' \cite{binwalk}.
    \item \textbf{Step 4:} By using the search functionality provided by Hex Editor Neo \cite{hexeditor_link}, we have discovered that the AP uses a proprietary file system named MiniFS in this work.
    \item \textbf{Step 5:} In the editor (see Figure~\ref{fig:raw_no_mark}), we have discovered a structure of binary data different from the one presented in Section~\ref{sec:mfs_background}. The new version of the MiniFS system is described in Section~\ref{sec:minifs_struct}. Since it was impossible to use existing tools \cite{GIToldMINIFS}, we extracted the file system as detailed in Section~\ref{sec:eva}.
\end{itemize}

Note that due to limited space, Step 6 is left for future research publications.



\section{MiniFS File System}
\label{sec:minifs_struct}
This section describes the MiniFS structure; how files and file attributes are stored. We gained this information through several iterations of reverse engineering.

\begin{figure}[b]
    \centering
    \includegraphics[width=0.2\textwidth]{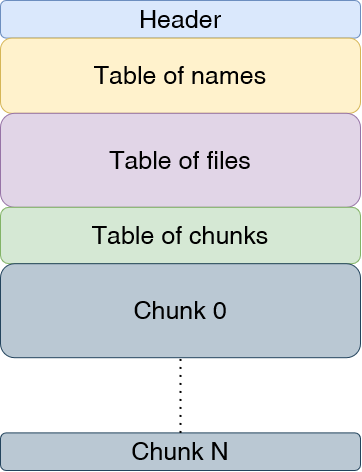}
    \caption{MiniFS structure.}
    \label{fig:minifs_struct}
\end{figure}

The file system begins with a header containing a special marker used to locate the file system in the raw binary image, along with several essential fields for further processing. The next element is the Table of Names, which stores full paths and file names, followed by the Table of Files. Next, there is a table containing information about each data chunk, followed by a set of these chunks (data is compressed using the~\ac{LZMA} algorithm). The graphical representation of the file system structure is depicted in Figure~\ref{fig:minifs_struct}, with detailed descriptions of each region provided in Sections~\ref{sec:fmt_header}~--~\ref{sec:fmt_data}. Note that all fields in the utility tables are presented in big-endian format and have a length of $4$ bytes.

\subsection{Header}
\label{sec:fmt_header}

\begin{figure}[h!]
    \centering
    \includegraphics[width=\columnwidth]{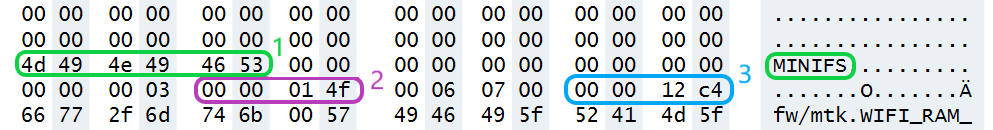}
    \caption{MiniFS's header.}
    \label{fig:header_decode}
\end{figure}

The header (see Figure~\ref{fig:header_decode}) is a set of binary fields with a start marker in the form of an ASCII string \textbf{"MINIFS"} and has a fixed size of 32 bytes. We have figured out the purpose of some fields, but not all of them. The start address of the magic string is the base address for decoding data. In Figure~\ref{fig:header_decode}, fields of the record's data are highlighted with different colors. If the data can be represented as an ASCII string, the representation area on the right is highlighted with the same color. Numbers above the fields correspond to the Table~\ref{tab:header_table}. This principle is applicable to all the images presented below. 
{Note that the header contains several unidentified fields requiring further investigation. Hypothetically, one of the fields may contain the file system version. The other field having an unknown purpose, in our device, contains a value coinciding with the size of the i) first file and ii) unpacked first chunk.}

\begin{table}[h!]
\centering
\caption{Structure of the header}
\label{tab:header_table}
\begin{tabular}{cccll}
\hline
$N$ & \textbf{Offset} & \textbf{Length} & \textbf{Description} & \textbf{Value}                                                                     \\ \hline
$1$ & \texttt{0x00}   & $6$      & Magic number& MINIFS                                                              \\
$2$ & \texttt{0x14}   & $4$      & Number of files& \texttt{0x14F}\\&&&&\textrightarrow 335 files    \\
$3$ & \texttt{0x1C}   & $4$      & Table of Names Size& \texttt{0x12c4}\\&&&&\textrightarrow $4,804$ bytes
\end{tabular}%
\end{table}

\subsection{Table of Names}
\label{sec:fmt_names}

\begin{figure}[h!]
    \centering
    \includegraphics[width=0.48\textwidth]{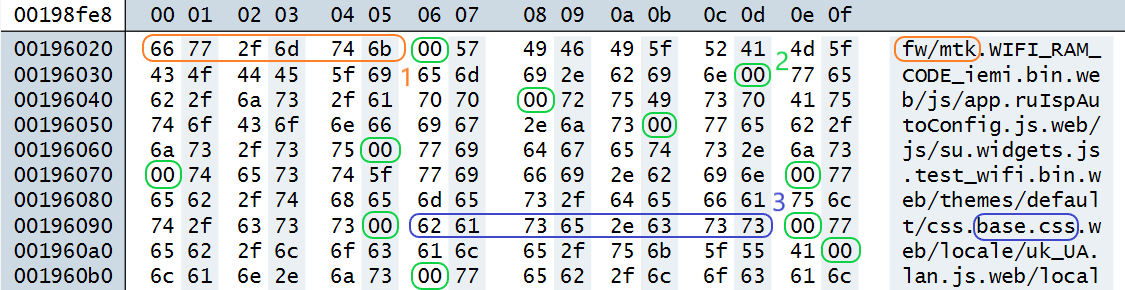}
    \caption{MiniFS's Table of Names.}
    \label{fig:names_decode}
\end{figure}

The Table of Names (ToN) consists of a set of ASCII symbols with a delimiter in the form of 0x00 between the fields see in Figure~\ref{fig:names_decode} and Table~\ref{tab:names_table}. The name table does not have a fixed length for its elements. This is a mixture of full paths to files and file names, done to save space, as multiple files may be located under the same path. To create the full path for a specific file, you need to take its full path, add a slash symbol '/', and append the file name. The table of names is located in the address: $base + 32$ $bytes$.

\begin{table}[h!]
\centering
\caption{Table of Names structure}
\label{tab:names_table}
\begin{tabular}{ccll}
\hline
N   & Length    & Description       & Value    \\ \hline
$1$ &           & A full path name. & fw/mtk   \\
$2$ & $1$       & Field delimiter.  & \texttt{0x00}     \\
$3$ &           & A file name.      & base.css
\end{tabular}%
\end{table}
\subsection{Table of Files}
\label{sec:fmt_files}

\begin{figure}[h!]
    \centering
    \includegraphics[width=0.48\textwidth]{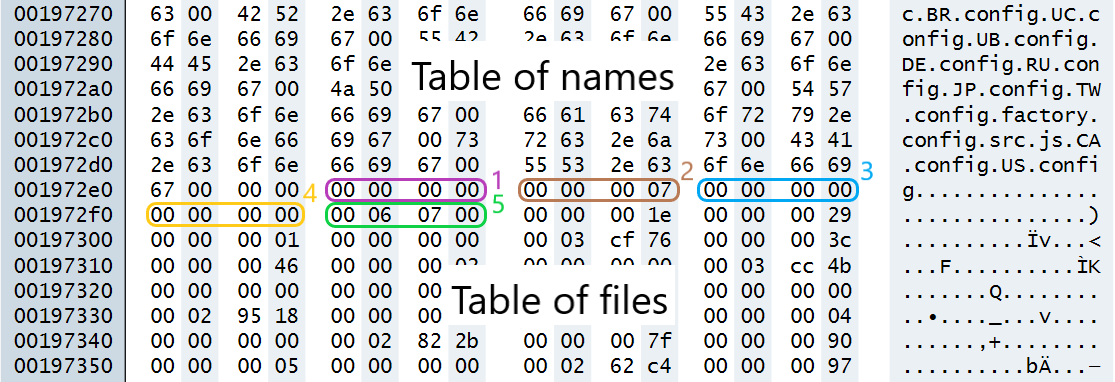}
    \caption{MiniFS's Table of Files.}
    \label{fig:files_decode}
\end{figure}

The Table of Files consists of binary format records, each with a length of 20 bytes for every file. The table includes entries for all files in the file system. It is presented in Figure~\ref{fig:files_decode}. It contains information on how to correctly assemble the file name with the full path, in which chunk and at what offset the file data is located, and what size it is. The table of files locate in address: $base + 32$ $bytes + size(table$ $of$ $names)$ $ bytes$\textrightarrow $base + 32$ $bytes + 4804$ $bytes$.

Table~\ref{tab:files_table} shows the structure of a record for a file in the file table with an example for the first file.

The initial chunk is numbered 0. To determine the total number of chunks in the binary image, one should examine the last file description and increment the chunk number by 1. In this case, it is $0$x$29$\textrightarrow $41 + 1$\textrightarrow $42$ chunks.
\begin{table}[h!]
\caption{Table of Files structure}
\label{tab:files_table}
\begin{tabular}{cccll}
\hline
N & Offset & Length & Description&Value                                                      \\ \hline
1 & 0x00   & 4      & ToN offset for path& 0 bytes                  \\
2 & 0x04   & 4      & ToN offset for file name& 7 bytes             \\
3 & 0x08   & 4      & File location chunk number& 0                          \\
4 & 0x0C   & 4      & Offset within the chunk& 0 bytes                   \\
5 & 0x10   & 4      & File size& 0x60700\\
&&&&\textrightarrow 395008\\&&&& bytes
\end{tabular}%
\end{table}

\subsection{Table of Chunks}
\label{sec:fmt_chunks}

\begin{figure}[h]
    \centering
    \includegraphics[width=0.48\textwidth]{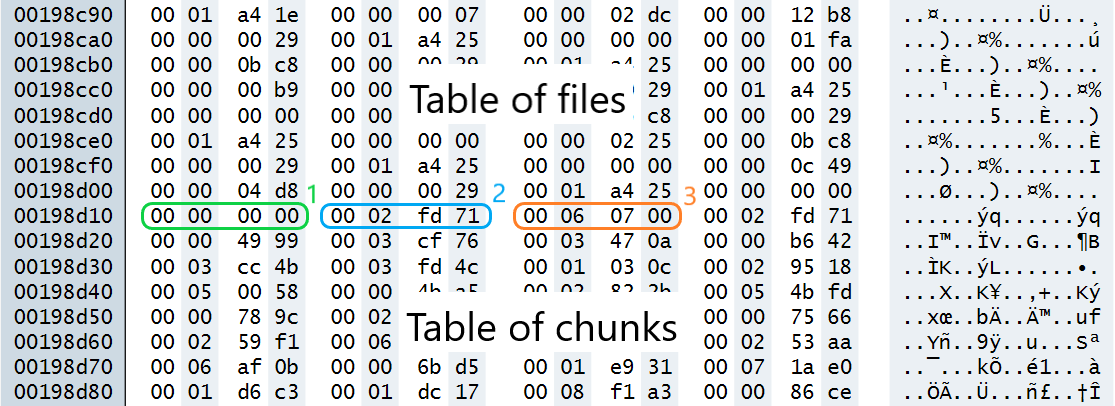}
    \caption{MiniFS's Table of Chunks.}
    \label{fig:chunks_decode}
\end{figure}

The Table of Chunks is a set of binary format records with a length of 12 bytes per chunk. The table contains entries for all chunks. In our case, its length is 42 fields. It is presented in Figure~\ref{fig:chunks_decode}. The information includes the byte offset of the chunk, the size of the chunk, and the data size after decompression. The Table of Chunks is located in address: $base + 32$ $bytes + size(table$ $of$ $names) + 20 \times files$\textrightarrow $base + 32$ $bytes + 4804$ $bytes + 6700$ $bytes$.
The address of the first chunk is used as the base address for the byte offset of raw chunks. It is $base + 32$ $bytes + size(table$ $of$ $names) + 20 \times files + 12 \times chunks$\textrightarrow $base + 32$ $bytes + 4804$ $bytes + 6700$ $bytes + 504$ $bytes$.

\begin{table}[h!]
\caption{Table of Chunks structure}
\label{tab:chunks_table}
\begin{tabular}{cccll}
\hline
№ & Offset & Length & Description &Value                                                           \\ \hline
1 & 0x00   & 4      & Chunk offset& 0 bytes                                             \\
2 & 0x04   & 4      & Chunk size& 0x2fd71\\&&&&\textrightarrow 195953 bytes         \\
3 & 0x08   & 4      & Decompressed size& 0x60700\\&&&&\textrightarrow 395008
\end{tabular}%
\end{table}
Table~\ref{tab:chunks_table} shows the structure of a record for a chunk in the chunk table with an example for the first chunk.

\subsection{Raw Data Chunks}
\label{sec:fmt_data}

\begin{figure}[t]
    \centering
    \includegraphics[width=0.48\textwidth]{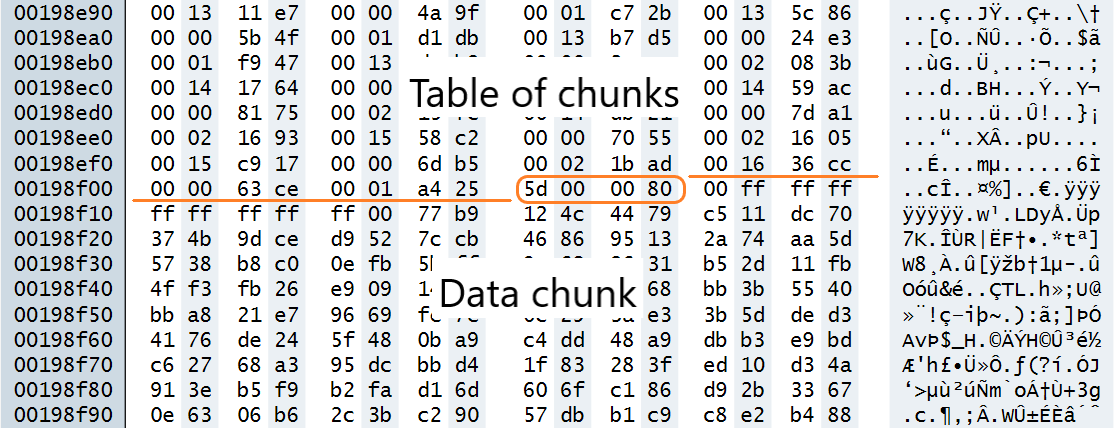}
    \caption{MiniFS's data chunk.}
    \label{fig:data_chunk}
\end{figure}

Raw data chunks consist of a set of binary data from one or multiple files compressed using \ac{LZMA}.
First raw chunk depicted in Figure~\ref{fig:data_chunk}. Chunks in the shared binary array follow one another without separators and can have varying sizes. At the beginning of each chunk, there is the LZMA configuration word 0x5D000080 (highlighted in the figure). This value is the same for all chunks. To verify that all chunks are correctly extracted from the binary file, the first 4 bytes can be checked for compliance with the configuration word.

\subsection{Comparison of the MiniFS versions}
Significant differences between the versions described in our research and those documented in existing literature (see Section~\ref{sec:mfs_background}) have been noted. Here we aim to highlight the discovered similarities and differences:
\begin{itemize}
    \item \textbf{File System Identification}: The identification method of the MiniFS file system within flash memory consistently relies on the magic string "MINIFS".
    
    \item \textbf{Header Format Variations}: A notable divergence in the header format of MiniFS versions has been observed. Previous sources describe the header as including a marker followed by an unidentified 4-byte field at offset 0x10. In contrast, our findings reveal a header composition that encompasses a marker, file count, and name table length, alongside unidentified 4-byte fields at offsets 0x10 and 0x18. This variation indicates an evolution in the header structure, potentially offering enhanced file system metadata.
    
    \item \textbf{File Property Encoding}: MiniFS in \cite{minifs_article_1,minifs_article_2} utilized an 88-byte fixed-size field for encoding file properties, integrating file size, offset, and a fixed-size 80-byte name and path field. Our analysis, however, indicates a separation of name and path into distinct fields within a single name table. This modification suggests an optimization in memory usage, facilitating more efficient file management.
    
    \item \textbf{File Storage Mechanism}: Contrary to the single-file-per-chunk approach previously documented, the current version supports multiple files within a single chunk. This enhancement likely reflects an optimization in storage utilization and file organization.
    
    \item \textbf{LZMA Configuration Word}: The LZMA configuration word at the beginning of a chunk has undergone a change, moving from 0x5A000080 in earlier sources to 0x5D000080 in our observations.
\end{itemize}

These comparative insights into the MiniFS versions not only illuminate the file system's development over time but also highlight the nuanced design choices aimed at optimizing performance and storage efficiency.

\section{MiniFS file system Extraction}
\label{sec:eva}

\subsection{Single File}
To decode a file, we extract information about the file from the  Table of Files, identify the chunk number in which it is located, and utilize the chunk table to determine the memory offset and size of that specific chunk. After reading and unpacking data using LZMA, we copy the necessary segment based on the details regarding offset and size contained in the Table of files. The result is saved at the full path, constructed using information about the file's path and name. This process enables the restoration of each file in the file system.

\subsection{Complete File System}
\subsubsection{Data sections identification}
\label{sec:minifs_local}

\begin{figure}[h]
    \centering
    \includegraphics[width=1\columnwidth]{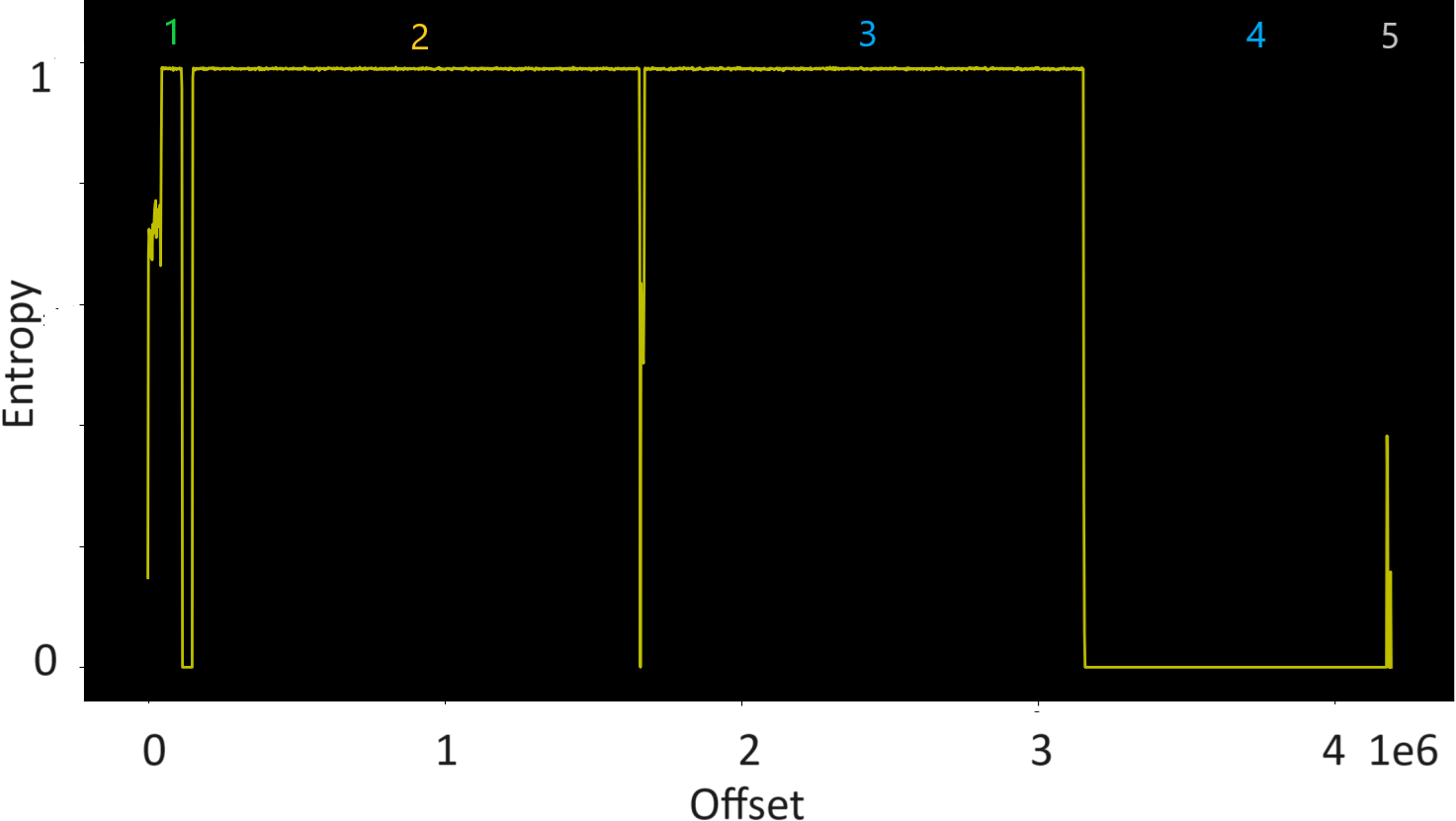}
    \caption{Flash memory binary image entropy.}
    \label{fig:image_entropy}
\end{figure}

We conducted entropy analysis of the entire binary image, and the image is illustrated in Figure~\ref{fig:image_entropy}. Five primary data sections can be distinguished within the raw binary image. In Figure~\ref{fig:image_entropy}, the identified sections are marked with numbers corresponding to the Table~\ref{tab:regions_table}.

\begin{table}[h]
\centering
\caption{Identified data sections}
\label{tab:regions_table}
\resizebox{5cm}{!}{%
\begin{tabular}{cl}
\hline
N & Description                         \\ \hline
$1$ & Bootloader.                         \\
$2$ & VxWorks \ac{RTOS}. \\
$3$ & MiniFS file system.                 \\
$4$ & Empty space.                        \\
$5$ & Configuration of the device.       
\end{tabular}%
}
\end{table}

\subsubsection{File system extraction flow}
\label{sec:extraction_all}
To extract all files from this file system using a binary image of the flash memory, one needs to follow a few simple steps:

\begin{enumerate}
    \item Find the magic number in a binary image and determine i) the number of files and ii) the table of names size.
    \item Extract Table of Files.
    \item Extract Table of Chunks.
    \item Extract chunks and unpack them by \ac{LZMA}.
    \item Build a working tree of folders using information about paths from the Table of Files.
    \item Construct the complete path and, utilizing details such as chunk number, offset, and size, segment it into a new file at the designated location.
\end{enumerate}

\subsection{Results}
Following the conducted research, a Python code has been developed to decode all files encoded in MiniFS. To assess the accuracy of the decoding, we performed an initial analysis of the data. The identified files include:

\begin{enumerate}
    \item CSS files: 19
    \item HTML pages: 70
    \item JavaScript files: 156
    \item Images: 8
    \item Various configuration files: 20
    \item Keys and certificates: 3
    \item Binary files for MTK7629: 8
    \item Utility configuration files for WiFi on MTK7629: 16
\end{enumerate}

\section{Conclusion and Future Work}
\label{sec:conclusion}

Through the reverse engineering of the MiniFS file system utilized in TP-Link's AC1900 WiFi router, this study has significantly advanced our understanding of the previously underdocumented file system. The described approach to data extraction has not only clarified the structure and operation of MiniFS but also underscored the variety within file system implementations in embedded devices.

Preliminary investigation of the decoded content has uncovered several private keys (hypothetically, these keys might be uniform across the entire range of these routers). Moreover, we would like to emphasize the lack of cryptographic protection within MiniFS, coupled with a reliance on security-by-obscurity, presents significant vulnerabilities in an era where digital connectivity is widespread.

Moving forward, the insights gained from this study will guide our examination of other WiFi access points operating on VxWorks. This effort is crucial for developing a more comprehensive understanding of security measures across devices, aiming to fortify the digital infrastructure against potential cyber threats. By advancing the methodology for analyzing and securing embedded systems, this research contributes to the broader endeavor of safeguarding user privacy and data integrity in our interconnected world.

\begin{acks}
We express our sincere gratitude to the reviewers for their thorough analysis and valuable comments, which have significantly improved the quality of our paper. Their critical remarks and constructive suggestions helped us make important changes and enhance the accuracy and clarity of our work. Moreover, we would like to extend special thanks to our colleague Pietro. His careful reading, deep understanding of the topic, and constructive criticism allowed us to improve the article. Thank you, Pietro, for your contribution and support, which made this work much better.
\end{acks}

\newpage
\bibliographystyle{ACM-Reference-Format}
\bibliography{references}

\end{document}